# The importance of categorization of nanomaterials for environmental risk assessment

Willie Peijnenburg[a,b]

[a] National Institute of Public Health and the Environment, Bilthoven, The Netherlands
[b] Institute of Environmental Sciences (CML), Leiden University, Leiden, The Netherlands

Email: willie.peijnenburg@rivm.nl

ORCID iD: 0000-0003-2958-9149

**Abstract**
Nanotechnology is a so-called key-emerging technology that opens a new world of technological innovation. The novelty of engineered nanomaterials (ENMs) raises concern over their possible adverse effect to man and the environment. Thereupon, risk assessors are challenged with ever decreasing times-to-market of nano-enabled products. Combined with the perception that it is impossible to extensively test all new nanoforms, there is growing awareness that alternative assessment approaches need to be developed and validated to enable efficient and transparent risk assessment of ENMs. Associated with this awareness, there is the need to use existing data on similar ENMs as efficiently as possible, which highlights the need of developing alternative approaches to fate and hazard assessment like predictive modelling, grouping of ENMs, and read across of data towards similar ENMs. In this contribution, an overview is given of the current state of the art with regard to categorization of ENMs and the perspectives for implementation in future risk assessment. It is concluded that the qualitative approaches to grouping and categorization that have already been developed are to be substantiated, and additional quantification of the current sets of rules-of-thumb based approaches is a key priority for the near future. Most of all, the key question of what actually drives the fate and effects of (complex) particles is yet to be answered in enough detail, with a key role foreseen for the surface reactivity of particles as modulated by the chemical composition of the inner and outer core of particles. When it comes to environmental categorization of ENMs we currently are in a descriptive rather than in a predictive mode.

## Introduction

Nanotechnology is a rapidly evolving technology with the potential to revolutionize the modern world. Materials take on entirely new chemical and physical properties at the nanoscale. This opens up totally new possibilities for material scientists but also commits them to assure a safer production, handling, and use of these materials. The novel properties of engineered nanomaterials (ENMs) are not only reason for enthusiasm, but also a potential cause of human health and environmental hazards beyond that of corresponding materials at larger sizes. It is crucial for developers of nanotechnology to learn about the most important parameters governing the properties, behaviour, and toxicity of ENMs. Given the almost exponential growth of the field of nanotechnology and the fact that the time-to-market of new products is rapidly becoming shorter, it is pivotal for unhindered industry-driven development of ENMs that validated and scientifically justified predictive models and modelling techniques are available and in use that allow for accurate screening of potential adverse effects. For regulators, it is important that predictive models are available that allow assessment of 'similarity' between different ENMs or different forms of an ENM to support decision making on whether to accept risk assessment on the basis of a category approach, or demand a separate risk assessment on a case-by-case basis.

Manufacturing and functionalising of materials at the nanoscale leads to a whole array of ENMs varying not only in chemical composition, but also, for example, in size, morphology and surface characteristics. Apart from expected benefits, distinctive properties of ENMs may also affect human health and the environment. Risk assessment requires sufficient information for each ENM, but testing every unique ENM for their potential adverse effects would be highly resource demanding. More efficient ways to obtain risk information are needed, and this could be achieved by applying these categorization approaches like grouping and read-across to ENMs. Some of the scientific foundations for the application of categorization approaches to ENMs have been established in a number of conceptual schemes as developed in the EU-funded projects MARINA [1], NANoREG [2], ITS-NANO [3] and in the ECETOC Nano Task Force [4]. In addition, European regulatory bodies and related expert committees have provided recommendations on how to identify ENMs and apply grouping and read-across to ENMs of the same substance in the context of REACH [5-7]. One of the major conclusions of these activities is that future categorization strategies should be hypothesis-driven and must consider not only intrinsic properties and (eco)toxicological effects, but also extrinsic (system-dependent) descriptors of exposure, toxico-kinetics and environmental fate.

## Categorization of nanomaterials

When searching the internet, there are various ways of facilitating a search. The category of natural products can for instance be restricted to fruits and vegetables and subsequently be categorized according to colour, size, or even price. Whether such a categorization is useful depends on the needs and purpose of the user. Similarly for ENMs, the needs and purposes of the user should be clear as categorization just for the purpose of categorization is not relevant for any setting, and lacks relevance especially for regulatory and innovative settings. Categorization of ENMs can serve various purposes:

- **To facilitate targeted testing or targeted risk assessment.** If it is known that one or more aspects (e.g. a physicochemical property) of a material may inform exposure, fate, and kinetic behaviour or a specific hazard; this knowledge can be used to target information gathering and testing for risk assessment, or to highlight specific points of interest when assessing the risk. The latter may e.g. be relevant for a substance evaluation under REACH, where one may focus specifically on certain aspects such as human inhalation risks or hazards for the aquatic environment. Several similar materials sharing known exposure, fate, kinetic or hazard information may be seen as an initial group as well as a starting point for hypothesis formulation.
- **To fill data gaps in regulatory dossiers.** When a regulatory dossier on a chemical is submitted to a regulatory agency, it may be possible to provide the requested information by grouping chemicals based on similarity and by applying read-across, i.e. use information from other (groups of) similar chemicals to predict required information and fill data gaps. REACH is the regulatory framework that has the most advanced legislation with regard to grouping and read-across, as these options are specifically mentioned in the legal text as a means of fulfilling information requirements [8]. Other legal frameworks in the EU and international organisations such as the Organisation for Economic Co-operation and Development (OECD) apply or discuss grouping and read-across for chemicals and nanomaterials (e.g. [9, 10]).

- **To develop precautionary measures.** Based on the known information on exposure, fate, kinetic behaviour or hazard of similar materials, precautionary measures can be taken for a new material for which that information is not available, e.g. by reducing or preventing exposure.
- **To steer safe innovation/safe-by-design.** For a new material under development, information available on similar materials or relationships, for example, with physicochemical properties can provide an indication of potential issues with exposure, fate, kinetic behaviour, or hazard. This approach provides an opportunity to exploit this information to steer safe innovation and safe-by-design. Also, knowledge on the likelihood to use grouping and read-across later in the innovation process is relevant, as targeted testing and read-across approaches will likely reduce needed resources and be less time-consuming than case-by-case testing to satisfy regulatory information requirements to obtain market approval under a specific law.
- **To improve scientific understanding.** For example, modelling (e.g. quantitative structure-activity relationships, QSARs) of the behaviour of ENMs (fate/toxico-kinetic behaviour, effects) can lead to new insights in fate and effect-related material properties that can in turn lead to establishing new groups of ENMs and to new read-across options. When the scientific understanding increases, the possibilities of grouping of ENMs increase, and vice versa, identifying possibilities for grouping may increase scientific understanding. This scientific knowledge and understanding can be used in regulation, for targeted testing, safe-by-design, etc.

In practical terms, categorization involves treating groups of similar substances as a category. Missing data on endpoints or properties within a category are predicted by read-across from data-rich analogues within the category. The way similarity is defined within a group is essential to read-across. Unfortunately, there is no one single approach to define similarity whereas similarity is endpoint-dependent. Also, no formal rules or common practices exist for determining the validity of chemical categories. It is nevertheless obvious that justification of the scientific robustness of category-based data gap filling approaches is required before application of categorization. In general, there is a preference for the use of interpolation within categorization approaches as this gives rise to less uncertainty than in case of extrapolation. In risk assessment, the exception to this preference is where an extrapolation from one substance to another leads to an equally severe or more severe hazard for the target substance. Although it may seem logical to assume that interpolation is subject to less uncertainty than extrapolation, *in reality, the degree of uncertainty is not due to the interpolation or extrapolation of data, but rather to the strength of the relationship forming the basis of the category/analogue approach itself.* This in turn is dependent on the size of the category and the amount and quality of the experimental data for the category members themselves. If the relationship underpinning the category is poorly defined, then interpolation or extrapolation can result in significant uncertainty.

Categorization of ENMs should provide a valuable means of filling data gaps essential for proper ENM risk assessment, including fate properties as well as hazardous effects. For the prediction of ENM properties on the basis of categorization and subsequent read-across of available data, three options can be foreseen: 1 – from bulk to all nanoforms; 2 – from bulk to specific nanoforms; 3 – from one or more nanoforms to one or more other nanoforms. In all cases, the nanoforms may be of either the same chemical identity or of the same chemical identity but with differences in physicochemical characteristics, including differences in the surface composition and surface chemistry. The key properties that characterize an ENM are exemplified in Figure 1, distinguishing four property classes that in turn might be categorized as indicating 'what they are' (chemical and physical identity), 'where they go', and what they do.

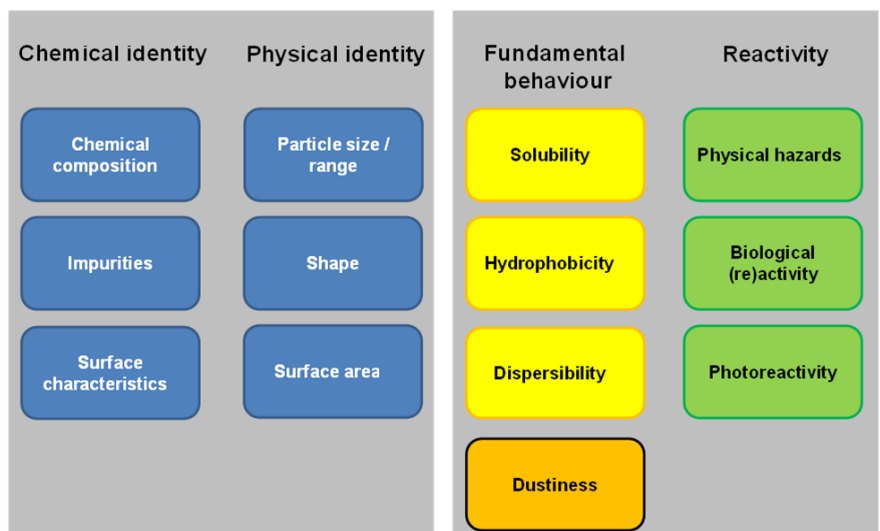

**Figure 1:** Schematic overview of the key properties that characterize an ENM

Arts et al [4] were the first to propose a framework for grouping and testing of ENMs. Fulfilling the requirement identified above on needs and purpose of categorization, the framework was proposed with the clear objective of distinguishing groups of metal oxides and metal sulphates with regard to *in vivo* inhalation toxicity. Based on the intrinsic material properties depicted in Figure 1, system dependent properties like dissolution, dispersability, and surface reactivity, and information on effects of metal oxides and metal sulphates in a short-term rat inhalation study, four main groups of ENMs were distinguished:
1 – Soluble, non-biopersistent ENMs like ZnO and CuO for which the chemical composition is more important for hazard assessment than the as-produced nanostructure.
2 – Biopersistent and rigid high aspect ratio ENMs for which there are concerns related to their asbestos-like hazards.
3 – Passive, biopersistent, non-fibrous ENMs like $BaSO_4$ that do not possess a toxic potential.
4 – Active, biopersistent, non-fibrous ENMs like $CeO_2$ and $TiO_2$ that are potentially hazardous.

**Driving forces for environmental categorization of nanomaterials**
It is likely that categorization of ENMs with regard to environmental hazards is likely to yield a framework that is in general terms similar to the framework advocated by Arts et al [4]. As asbestos-like behaviour is irrelevant for the endpoints commonly considered in environmental risk assessment, it is obvious that the category of biopersistent and rigid high aspect ratio ENMs is not relevant for environmental categorization of ENMs. Until now no efforts have been undertaken to systematically develop a classification framework for the purpose of environmental risk assessment of ENMs. When developing such a framework, the *key question* that is the basis for categorization of ENM from an environmental point of view, is: *What drives fate and effects of ENMs?* In answering this question, several considerations are of relevance. First, it is to be realized that it is preferred for environmental categorization to take all life stages of the material into account, whilst explicitly considering all environmental impacts as commonly done within life cycle assessment (LCA). This is schematically illustrated in Figure 2.

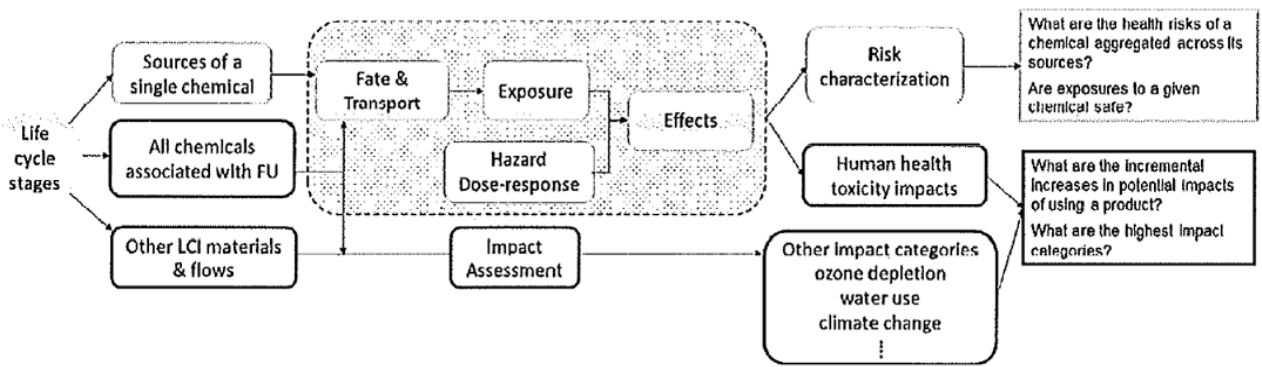

**Figure 2:** Schematic overview of the various assessment steps within LCA, including environmental risk assessment

The environmental impacts are calculated on the basis of the emissions into any of the environmental compartments during each of the life stages of the ENM. Given the perceived overall aim of environmental categorization of supporting risk reduction in order to minimize adverse effects induced by emissions of ENMs during any of the life stages, categorization can be applied with the purpose of:
1 – Reduction of exposure.
2 – Reduction of hazard as assessed on the basis of dose-response relationships typically derived in a laboratory setting.

Current research on exposure assessment of ENMs has shown that the fate of ENMs is usually determined by the physicochemical characteristics of the particles and the environmental conditions and can best be modelled using kinetic models instead of equilibrium-based models commonly applicable for dissolved organic compounds [11-13]. Modeling exercises have shown that in general, only a limited number of key processes drive the actual exposure of biota to ENMs. These processes include sorption of biomolecules (organic carbon), transformation, and heteroaggregation. Examples of classification approaches for these key processes are not yet available. For the case of sorption of biomolecules to ENMs, particle size, particle morphology, and surface charge are the predominant drivers. Basically, similar to the findings of Arts et al [4], in case of transformation there are sound perspectives of defining categories of ENMs for which the combination of intrinsic reactivity and environmental conditions induces high, medium, or low reactivity. In case of highly reactive ENMs the focus of subsequent hazard assessment should be restricted to the transformation products instead of being on the pristine starting materials, whereas in the opposite case of low reactivity focus should be on the hazards of the particles themselves. The key challenge in this respect will be to define cut-off limits for the kinetics of transformation, in a first-tier approach based on a realistic basis scenario regarding the composition of the environmental media of relevance.

An interesting approach of environmental categorization for heteroaggregation has been developed by Meesters [14]. Applying the nano-specific fate model Simplebox4Nano [15], it was shown that attachment efficiency ($\alpha$) can be used as the sole factor for quantifying the faction of (bio)persistent nanoparticles in the water freely available for interaction with biota. In this specific case, two categories can be distinguished on the basis of a cut-off value for $\alpha$ of $10^{-4}$. As illustrated in Figure 3, particles for which $\alpha$ exceeds this cut-off value are likely to heteroaggregate with natural colloids or attach to natural coarse particles. Subsequent sedimentation implies that risk assessment of these particles should focus on the sediment compartment. Particles for which $\alpha$ is below the cut–off value of $10^{-4}$ will reside in the water phase and will govern the effective exposure of aquatic organisms.

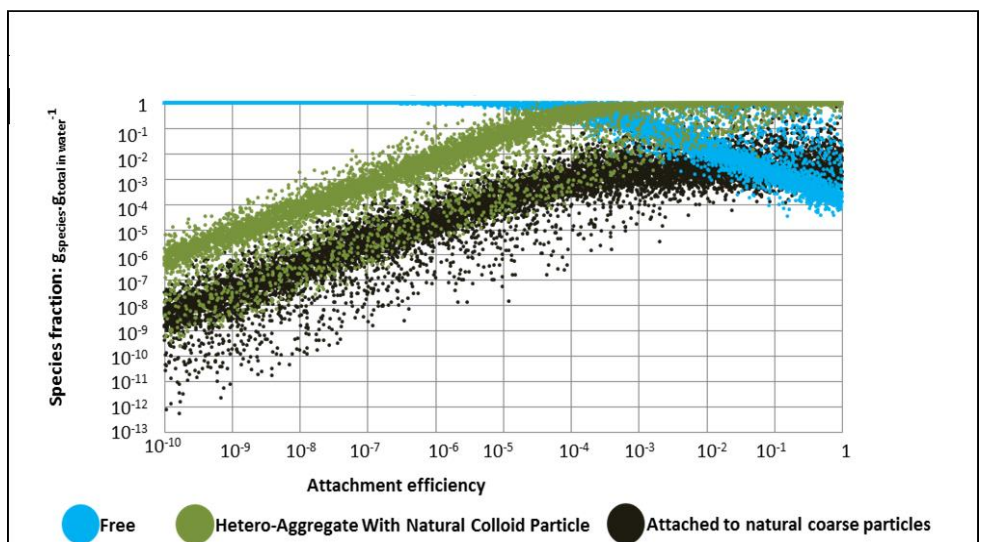

**Figure 3**: Simulation of the distribution of ENMs in the water column as arranged by attachment efficiency of the particles

In summary, this implies that only a limited number of particle properties are essential for classification of ENMs on the basis of their fate properties, whereas these properties can be classified as either extrinsic (transformation rate, attachment efficiency, and surface charge) or intrinsic (particle size, particle morphology).

Directly linked to the processes that determine the effective exposure concentrations of ENMs to biota, tools, methods, and insights are available for the purpose of ENM categorization to facilitate hazard assessment and hazard reduction. Until now, none of them have yet crystallized in a broadly applicable environmental categorization framework. The overarching challenge of developing such a framework may first of all be triggered by the wealth of scattered information on the factors affecting uptake and adverse effects of ENMs. It is for instance well-established that uptake of ENMs across epithelial membranes is dictated (among other factors) by size, shape and surface charge [16]. While size has been shown to influence uptake and biodistribution in zebrafish embryos [17, 18], the impact of different nano-shapes on biodistribution is less investigated. Particle shape can be an important factor for cellular uptake, circulation kinetics within the organism, and biodistribution of suspended particles [19]. In general, small, elongated colloidal particles are more easily taken up by cells than large and flat individual particles [20]. This same tendency was found for the endpoint of biodistribution, as in the case of gold ENMs nanorods distributed throughout tumor tissues, whereas spheres and discs were located only at the surface of tumor cells [21]. Moreover, the length of rods was found to determine uptake and internal distribution: short rods were taken up faster and were trapped in the liver, while longer rods showed lower uptake efficiency and were trapped in the spleen of mice [22-24]. Additionally, sharp gold nanostars can pierce the membranes of endosomes and escape to the cytoplasm regardless of their surface chemistry, size or composition [23, 25].

Size, morphology, and chemical composition are amongst the key factors modulating particle toxicity. As exemplified in Figure 4, the toxicity of rod-shaped particles is in general lower than the toxicity of differently shaped particles whereas toxicity increases upon decreasing particle size, offering opportunities for future systematic categorization of ENMs. In a quantitative sense, it was shown by Hua et al [26] that the ratio of particle-volume:particle-diameter is a superior dose descriptor to replace the conventional dose metrics of mass as commonly used for expression of toxicity of soluble chemicals.

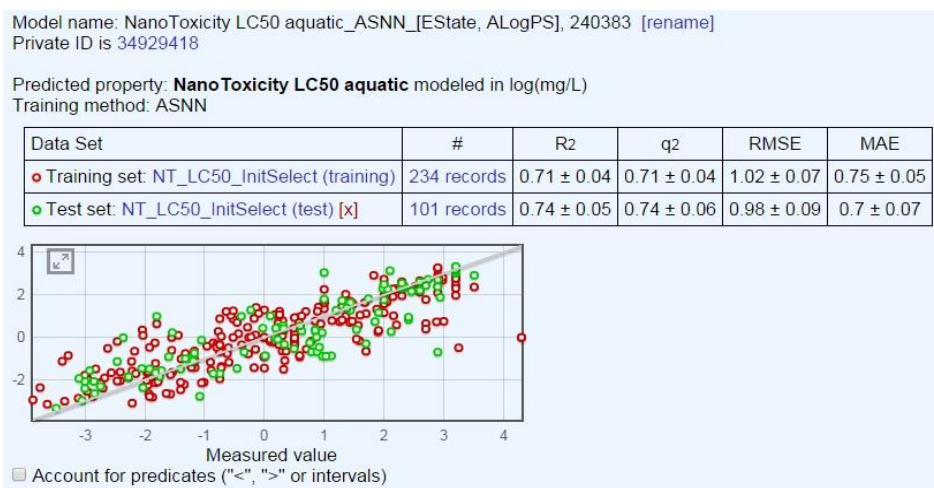

**Figure 4**: Impact of chemical composition (A: Ag, B: ZnO, C: Cu, D: Pb), size and particle morphology on toxicity of ENMs to micro-organisms

*In silico* methods like QSAR and grouping and read-across have been used for several decades to gain efficiency in regulatory hazard assessment of chemical substances in general and to improve animal welfare. Subsequently, guidance was developed for the implementation of these methods in regulation. OECD published, for instance, its first guidance on grouping of chemicals in 2007 [27] whereas ECHA published guidance on grouping of chemicals in 2008 [28] and the read-across assessment framework was updated in 2017 [29]. Neither of these documents mentions classification approaches for ENMs whereas OECD actually concluded in the second edition of its guidance on grouping of chemicals that development of guidance specifically for ENMs is premature [9]. Current efforts are directed towards development of ENM-specific QSARs, as reviewed by Chen et al [30]. An example of a generic ENM-specific QSAR is given in Figure 5. Apart from QSARs for endpoints that are relevant from a regulatory point of view, predictive models for nanomaterial hazard categorization have also received attention [31]. Unfortunately, these models have not yet reached sufficient maturity to allow for implementation in for instance risk assessment.

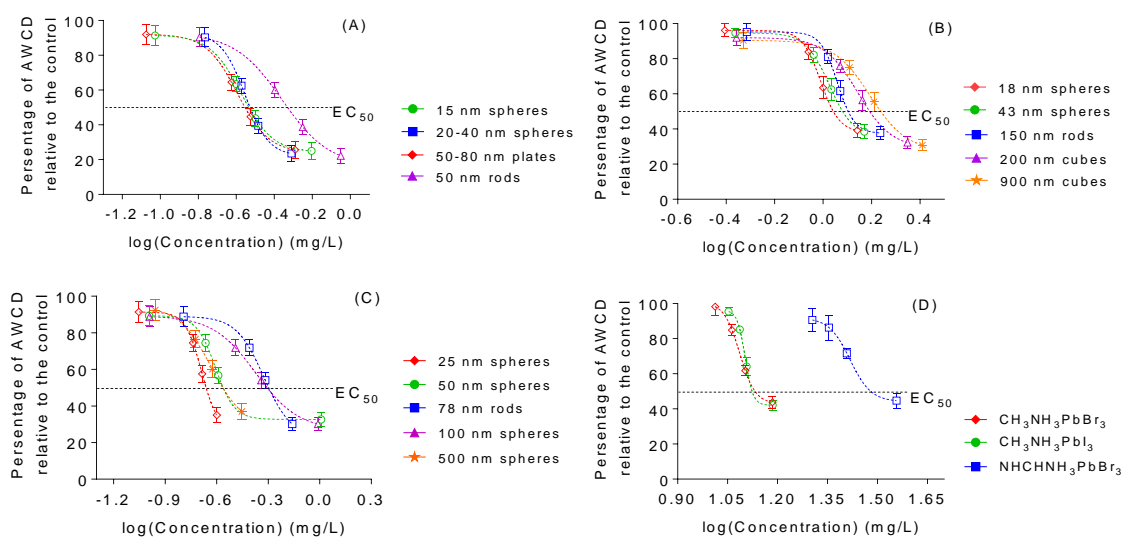

**Figure 5:** Example of a comparison of predicted (y-axis) and experimental aquatic LC50 values for a mixed set of ENMs as based on a dataset of 234 structurally different ENMs

## Conclusions

It is to be acknowledged that reduction of testing needs and efficient use of available data are the key drivers for environmental categorization of ENMs. Successful development, quantification, and validation of category approaches will increase the efficiency of risk assessment whilst respecting the principles of Replacement, Reduction and Refinement of animal testing. Broadly applicable predictive models for quantification of the key properties driving fate and effects of ENMs are currently in their early stage of development even though a number of models have successfully been generated. Fortunately, various qualitative approaches to grouping and categorization have been developed. Yet, these approaches need to be substantiated and additional quantification of the current sets of rules-of-thumb based approaches is a key priority for the near future. Most of all, it is to be concluded that the key question of what actually drives the fate and effects of (complex) particles is yet to be answered in more detail. Most likely, a key role is played in this respect by the surface reactivity of the particles as modulated by the chemical composition of the outer core, the dynamics of the outer core in terms of interactions with its surroundings, the chemical composition of the inner core, and the number of available atoms on the particle surface, as well by other hitherto unexploited properties. Although this might seem to be a long way to go, experiences in the past have learned that various shortcuts are quite possible to speed up the process of efficient environmental risk assessment of ENMs. When it comes to environmental categorization of ENMs, we currently are in a descriptive rather than in a predictive mode.

## Acknowledgements

This article is one of a collection of articles about the categorization of nanomaterials, generated by research and workshop discussions under the FutureNanoNeeds project funded by the European Union Seventh Framework Programme (Grant Agreement No 604602). For an overview and references to other articles in this collection, see *The Nature of Complexity in the Biology of the Engineered Nanoscale Using Categorization as a Tool for Intelligent Development* by Kenneth A. Dawson.

Author declares there is no conflict of interest.